\documentclass[fp,twocolumn]{jpsj3-mod}
\usepackage{txfonts}

\title{Elastic Anomaly of Thin Neon Film}

\author{Takahiko Makiuchi, Katsuyuki Yamashita, Michihiro Tagai, Yusuke Nago, and Keiya Shirahama}
\inst{Department of Physics, Keio University, Yokohama 223-8522, Japan} 

\abst{
Adsorbed molecular films provide two-dimensional systems that show various emergent phenomena that are not observed in bulk counterparts. 
We have measured the elasticity of thin neon films adsorbed on porous glass down to 1 K by the torsional oscillator technique. 
The shear modulus of a neon film anomalously increases at low temperatures with excess dissipation. 
This behavior indicates a crossover from a soft (fluidlike) state at high temperatures to a stiff (solidlike) state at low temperatures. 
The temperature dependence of the anomaly is qualitatively similar to that of the elastic anomaly of helium films found in our recent study.
The dissipation peak temperature, however, becomes constant at about 5 K, contrary to the case of helium, in which it decreases to 0 K at a critical coverage of a quantum phase transition between a gapped localized phase and a mobile (superfluid) phase. 
It is concluded that neon films behave as a classical system that does not show a quantum phase transition or superfluidity, although the films may be strongly supercooled to temperatures much lower than the bulk triple point, 24.6 K. 
Our results suggest that the elastic anomaly is a universal phenomenon of atomic or molecular films adsorbed on disordered substrates. 
 }

\begin{document}
\maketitle

\section{Introduction}
Adsorption is a ubiquitous phenomenon in everyday life, but understanding and controlling adsorption is still a challenging issue in both basic science\cite{TaoRappe} and modern technologies such as hydrogen storage\cite{BroomBook} and gas sensors\cite{Chen2013}. 
In condensed matter physics, adsorption of molecules on a solid surface has been utilized to produce two-dimensional (2D) films that show a number of emergent phenomena that are not observed in bulk materials. 
Molecules and noble gases (e.g., N$_2$ and Ar) form solid films at low temperatures, but 
helium is an important exception. 
It provides various phases other than the solid phase because of its quantum nature, namely, large zero-point fluctuation and weak interatomic potential. 
Both bosonic $^4$He and fermionic $^3$He exhibit various quantum phases such as superfluid, quantum solid, Mott insulator, and Fermi liquid phases, between which the transformation is by quantum phase transition (QPT)\cite{Casey2003,Neumann2007,Nyeki2017,Makiuchi}. 

There are other adsorbed films that may show quantum effects: hydrogen (H$_2$, HD, and D$_2$) and neon ($^{20}$Ne). 
Bulk hydrogen is a quantum solid where the amplitude of zero-point motion is comparable to the lattice constant \cite{VanKranendonkBook}. 
Neon is the second lightest noble gas and has the third largest quantum parameter, i.e. the ratio of the zero-point energy to the potential depth of pair interaction, following helium and hydrogen\cite{Nosanow1977}. 
Bulk solid neon is barely a quantum solid in the sense that the nearest-neighbor distance is slightly larger than the distance of the pair potential minimum because of 
zero-point motion\cite{BruchBook}. 
Therefore, neon may be used to investigate intermediate properties between quantum and classical.

In this paper, we report an elastic anomaly found in thin neon films adsorbed on a porous glass (PG) substrate. 
We have recently found that the elasticity of helium films on PG shows an anomalous behavior\cite{Makiuchi}.
It is known that $^4$He films on a disordered substrate such as glass show a QPT between a localized solid and superfluid when the coverage 
$n$ exceeds the critical value $n_{\mathrm {c}}\sim 20\ \mu$mol/m$^2$ (1.5 atomic layers). 
We measured the temperature ($T$) dependence of the elastic constants of $^4$He and $^3$He films below 1.1 K using a torsional oscillator (TO) specially designed for elasticity measurement. 
As $T$ increases, the localized phase changes from a stiff state 
to a soft state in which He films lose 
almost all the stiffness.
This elastic anomaly was explained by thermal activation of helium atoms from localized states to extended states, which are separated by an energy gap $\Delta$ that decreases continuously to zero as $n$ approaches $n_{\mathrm {c}}$. 
The localized phase of $^4$He and $^3$He films have an identical ground state, which is gapped and compressible. 
The gapped ground state may be a sort of Mott insulator or Mott glass, both of which can be formed by quantum effect and strong correlation in 2D helium films. 

The elastic study showed that the localized phase originates from quantum properties intrinsic to helium. 
Thus, it is natural to investigate the possible quantum effect in ``less quantum'' hydrogen and neon films. 
In this paper, we focus on neon\cite{hydrogen}. 
Neon films have not been extensively studied.
 Past heat capacity studies suggested that submonolayer neon films on graphite have 
2D solid and gas phases, and 2D solid--gas, solid--liquid, and liquid--gas coexisting phases\cite{Huff1976,Rapp1981}. 
Since 
 these measurements were limited to rather high temperatures, quantum properties of neon films have not been clarified. 
No systematic studies have been carried out on neon films on disordered substrates.

\section{Experimental Procedure}
\subsection{Torsional oscillator}
\begin{figure}
 \centering
 \includegraphics[width=60mm]{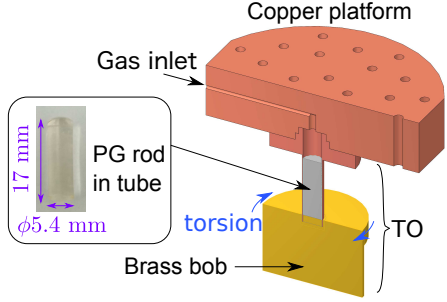}
 \caption{
 (Color online) Cross-sectional view of the TO.
 A porous glass (PG) rod is inside a thin BeCu tube.
 A brass bob is suspended from the rod.
 The TO is mounted on a massive copper platform with six screws (not shown).
 Neon gas was added through the gas inlet in the platform and the TO.
 }
 \label{fig:TO}
\end{figure}

We employed the same TO as that used in the studies on helium\cite{Makiuchi}.
The TO is schematically shown in Fig. \ref{fig:TO} with a photograph of a PG sample. 
The TO consists of a torsion rod containing a PG sample (5.4 mm in diameter, 17 mm long) and a brass bob, which also acts as an electrode for driving and detecting the torsional oscillation. 
The PG rod was covered by a thin BeCu tube of 0.2 mm thickness.
The TO was mounted on a massive copper platform (shown in Fig. \ref{fig:TO}) as a vibration isolator. 
The platform was set on a copper plate under a mixing chamber of a dilution refrigerator (DR) via a copper rod (not shown). 

The resonant frequency is $f = (1/2\pi)\sqrt{k/I}$, where $k$ is the torsion constant of the rod and $I$ is the moment of inertia of the bob. 
The change in $k$ upon the adsorption of neon is expressed by $k = k_0 + \delta k$. 
Hereafter, the subscript 0 denotes the quantity before the adsorption of neon.
The change in $f$ upon neon adsorption is derived from the above equation to the first order, 
\begin{equation}
\frac{2\delta f}{f_0} = \frac{\delta k}{k_0},
\end{equation}
where $f = f_0 + \delta f$ and $f_0\simeq 859$ Hz is a constant. 
In the measurement, the conditions $\delta f \ll f_0$ and $\delta k \ll k_0$ hold.
The contributions of a change in the shear modulus of the rod $G_\mathrm{g} = G_\mathrm{g0} + \delta G_\mathrm{g}$ and a change in the density of the PG (by adsorbing neon in the pores) $\rho_\mathrm{g} = \rho_\mathrm{g0} + \delta \rho_\mathrm{g}$ are calculated by finite element method simulation as follows\cite{Makiuchi} (subscript g denotes porous glass): 
\begin{equation}
 \frac{2\delta f}{f_0} = 0.197\frac{\delta G_\mathrm{g}}{G_\mathrm{g0}}
 \label{eq:dfG}
\end{equation}
and
\begin{equation}
 \frac{2\delta f}{f_0} = -1.33\times 10^{-4}\frac{\delta \rho_\mathrm{g}}{\rho_\mathrm{g0}}.
 \label{eq:dfrho}
\end{equation}
These linear relations hold in the experimental ranges. 
The contribution of the change in the shear modulus is three orders of magnitude larger than that of the change in density. 
This warrants that the observed frequency shift in this work 
 originates from the change in the elasticity of the neon film. 

The PG sample, commercially called Gelsil, has three-dimensionally connected nanopores. 
The pore diameter distribution measured from the N$_2$ isotherm has a peak at 3.9 nm. 
The surface area is 166 m$^2$ 
and the porosity is $p=0.54$. 
Using the molar volume of bulk solid neon at 1 K and 0.1 MPa, 
$v_\mathrm{m} = 13.39\ \mathrm{cm^3/mol}$ \cite{Rabinovich1988}, the monolayer coverage and the coverage corresponding to filling the pores are estimated to be 21.0 and 82.9 $\mathrm{\mu mol/m^2}$, respectively. 

\subsection{Neon film preparation}\label{sec:NePrep}
We used commercial neon gas with an impurity concentration of less than 10 ppm\cite{JapanFineProducts}. 
The natural abundances of isotopes $^{20}$Ne, $^{22}$Ne (bosons), and $^{21}$Ne (fermion) are 90.48, 9.25, and 0.27 \%, respectively\cite{Rosman1998}. 
Our gas sample should therefore have a ratio close to the natural one, although the ratio was not analyzed. 
The elasticity measurement was performed using a dilution refrigerator (DR), which can 
cool the TO on the mixing chamber to 10 mK. 
This measurement was limited to down to 1.2 K because the thermal connection was specialized to extend the measurement to over 20 K. 

A neon film with the desired coverage is prepared by adding a known amount of neon gas to the cryogenic substrate. 
The capillary for introducing neon into the DR was kept at a high temperature to prevent the solidification of neon in it.
For this purpose, the capillary was wound with a Manganin twisted wire heater and was not thermally anchored except near the mixing chamber.
The temperature of the capillary at the height of the still of the DR was measured and confirmed to be always higher than the temperature of the TO without using the heater. 

The neon gas was prepared at a standard volume of 84.0 cm$^3$ in a room-temperature gas-handling system (GHS), 
 and added into the cryogenic part 
by controlling the needle valve in the GHS. 
 Room-temperature pressures at the standard volume $P_1$ and after the needle valve $P_2$, and the pressure 
 near the gas inlet of the TO (see Fig. \ref{fig:TO}) $P_3$ were continuously monitored.
 We measured $P_3$ using a laboratory-made Kapton diaphragm capacitive pressure gauge\cite{CrowellPhD} at the mixing chamber.
The temperature and $P_1$ before and after the addition of gas were used to calculate the number of moles of neon added. 
The error in this value is about $\pm 0.03$\%, which originates from the resolution of room-temperature measurement. 
After the addition of neon gas, the temperature of the PG sample was kept above the bulk triple-point temperature of neon, 24.6 K, for more than 1 hour to distribute neon molecules as uniformly as possible. 
The sample was then slowly cooled 
 by the Joule--Thomson effect of circulating a $^3$He--$^4$He mixture in the DR. 
During cooling, we carefully kept $P_2$ and $P_3$ ($P_2\simeq P_3$) below the bulk sublimation curve of neon for as long as possible.
By performing this procedure, the neon molecules are adsorbed as uniformly as possible in the PG sample.
By elasticity measurement, we confirmed that no bulk solid grew outside the PG rod. 

We established the above-mentioned procedure as a result of unsuccessful uniform adsorptions of neon on the PG substrate. 
The first two coverages, which were intended to yield $n=$ 5 and 8 $\mathrm{\mu mol/m^2}$, were not properly prepared. 
Bulk solid neon was formed outside the PG and blocked the capillary, which we concluded from a rise in $P_2$ when the solid melted. 
We estimate the amount of solid from the volume and temperature in the capillary and $P_2$.
We assume that the uncertainty of the estimated amount of solid is 30\%, and we have the estimated coverages of $(4.0\pm 0.3)$ and $(4.5\pm 1.0)\ \mathrm{\mu mol/m^2}$, respectively. 

\section{Results and Analysis}
\subsection{Elastic anomaly} 
\begin{figure}
 \begin{center}
  \centering
  \includegraphics[width=80mm]{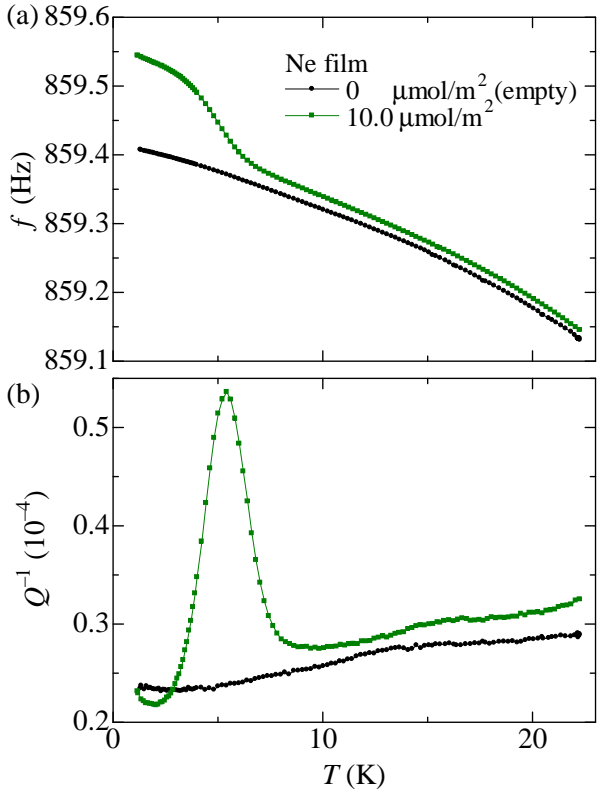}
  \caption{
  (Color online)
  Raw data of (a) resonant frequency $f$ and (b) energy dissipation $Q^{-1}$ at coverages $n = 0$ and $10.0\ \mathrm{\mu mol/m^2}$. 
  }
  \label{fig:rawdata}
 \end{center}
\end{figure}

In Fig. \ref{fig:rawdata}, we show the raw data of frequency $f$ and dissipation $Q^{-1}$ at $n = 0$ and $10.0\ \mathrm{\mu mol/m^2}$.
All the data shown in this paper were acquired during warming runs. 
The cooling trace was identical to the warming one. 
At $n = 0$, $f$ monotonically increases with decreasing $T$, while $Q^{-1}$ decreases down to 5 K, then shows a shallow minimum at about 3 K. 
We regard the $n = 0$ data as the background.  
When neon is adsorbed, $f$ and $Q^{-1}$ increases slightly at $T > 10$ K. 
Below 10 K, $f$ starts to deviate from the background $T$ dependence and increases rapidly at $T$ from 7 to 3 K. 
At lower $T$, $f$ returns to the same $T$ dependence as the background. 
The increase in $f$ accompanies a large excess dissipation, in which the peak is located at 5.4 K. 
We call these behaviors in $f(T)$ and $Q^{-1}(T)$ the elastic anomaly.

As shown in the previous section, there are two possible origins for the change in $f$, a change in the density of the PG sample, $\delta \rho_\mathrm{g}$, and a change in the elasticity of the PG, $\delta G_\mathrm{g}$. 
The frequency change determined from $\delta \rho_\mathrm{g}$ using Eq. (\ref{eq:dfrho}) is $-5$ mHz at $n=10\ \mathrm{\mu mol/m^2}$. 
Since this is much smaller in magnitude than the observed change in $f$, $+140$ mHz, and is negative, it is concluded that the increase in $f$ below 7 K 
originates from the increase in the elastic constant of the adsorbed neon film. 



\subsection{Coverage dependence of the anomaly} \label{sec:anomaly}
\begin{figure}
 \begin{center}
  \centering
  \includegraphics[width=80mm]{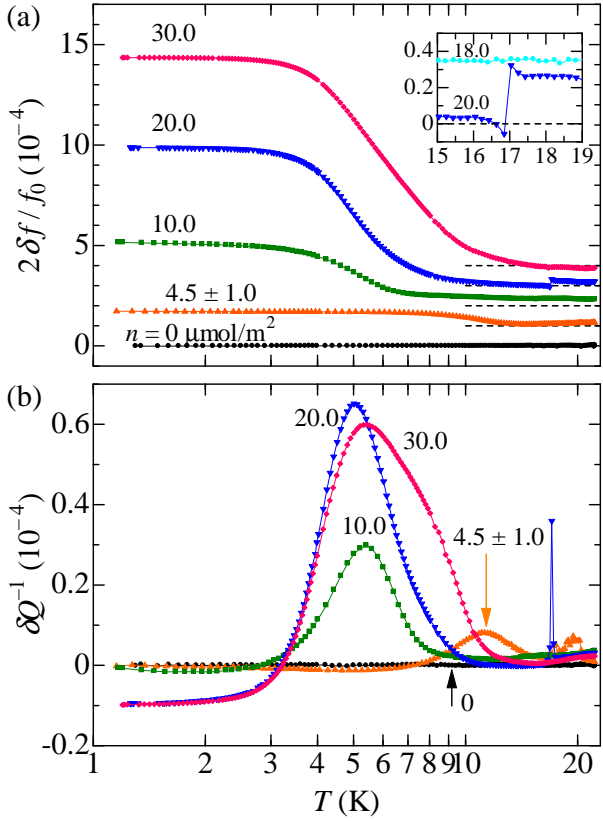}
  \caption{
  (Color online)
  Temperature dependence of (a) normalized resonant frequency shift $2\delta f/f_0$ and (b) excess dissipation $\delta Q^{-1}$ for selected coverages. The definitions of these quantities are given by Eqs. (\ref{eq:deltaf}) and (\ref{eq:deltaQinv}). 
The base frequency $f_0$ is taken to be 859.133 Hz for all coverages. 
  Data in (a) are vertically offset, indicated by dashed lines.
 The inset in (a) is an enlarged view of $2\delta f/f_0$ without and with anticrossing at $n=$ 18.0 and 20.0 $\mathrm{\mu mol/m^2}$, respectively.
  }
  \label{fig:dfdQinv}
 \end{center}
\end{figure}

To examine the elastic anomaly in detail, we extract the contribution of the neon film by subtracting the background from the raw data. 
We define the shifts in the frequency and dissipation from the background temperature dependence as 
\begin{eqnarray}
 &&\delta f(T) = f(T) - f_{\mathrm{B}}(T), \label{eq:deltaf}\\
 &&\delta Q^{-1}(T) = Q^{-1}(T) - Q^{-1}_{\mathrm{B}}(T),\label{eq:deltaQinv}
\end{eqnarray}
where subscript B denotes the background ($n = 0$). 
 These definitions ignore 
 the slight shifts in $f$ and $Q^{-1}$ at high temperatures 
 between finite-$n$ data and the background (e.g., the shifts in the data at $T > 10$ K in Fig. \ref{fig:rawdata}).
 The normalized frequency shift, which is given by $2\delta f(T)/f_0$, and $\delta Q^{-1}(T)$ defined in Eqs. (\ref{eq:deltaf}) and (\ref{eq:deltaQinv}) for four different coverages are shown in Fig. \ref{fig:dfdQinv}.
 The contributions from the slight shifts at high temperatures are small in both $2\delta f(T)/f_0$ and $\delta Q^{-1}(T)$ compared with the overall temperature dependence.

\begin{figure}
 \begin{center}
  \centering
  \includegraphics[width=80mm]{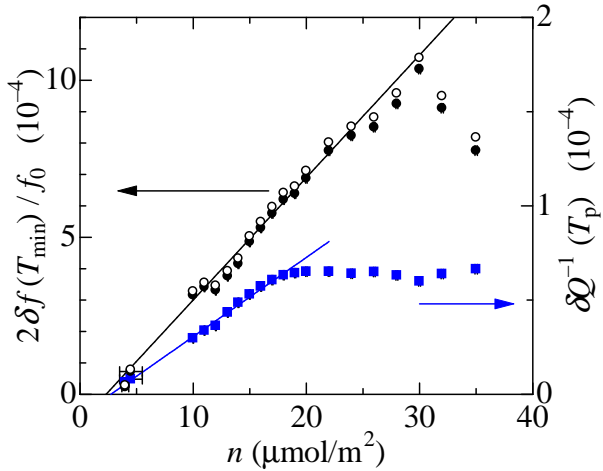}
  \caption{
  (Color online)
  Coverage dependence of the normalized resonant frequency shift at $T_\mathrm{min}\simeq 1.2$ K 
and the energy dissipation at $T_\mathrm{p}$. 
 Closed circles are $2\delta (T_\mathrm{min})/f_0$ determined using Eq. (\ref{eq:deltaf}), and open circles are corrected values (see the text).
  Solid lines are linear fittings, $y=a(n-n_\mathrm{offset})$, with slopes $a=(0.39\pm 0.01)$ and $(0.042\pm 0.001)\ \mathrm{m^2/\mu mol}$ for $2\delta f/f_0$ and $\delta Q^{-1}$, respectively. 
}
  \label{fig:n-dfdQ}
 \end{center}
\end{figure}

A number of characteristics are found from the data sets in Fig. \ref{fig:dfdQinv}. 
The normalized frequency shift $2\delta f/f_0$ is almost constant at high and low temperatures, and changes in the temperature range between 3 and 15 K. 
The peak of $\delta Q^{-1}$ is located at the inflection point of $\delta f$.
We define the temperature at the $\delta Q^{-1}$ peak as $T_\mathrm{p}$.
At $n=(4.5\pm 1.0)\ \mathrm{\mu mol/m^2}$, $T_\mathrm{p}\simeq 11$ K, whereas at higher coverages, $T_{\mathrm p}$ stays at about 5 K with a slight coverage dependence. 
The low-temperature limit of the frequency shift $2\delta f(T_\mathrm{min})/f_0$ with $T_\mathrm{min}\simeq 1.2\ \mathrm{K}$ increases monotonically as $n$ increases from 4.5 to $30\ \mathrm{\mu mol/m^2}$. 
This is shown in Fig. \ref{fig:n-dfdQ} and will be discussed later.
In the high-temperature region, $2\delta f(T)/f_0$ is almost equal to the background. 

\begin{figure}
 \begin{center}
  \centering
  \includegraphics[width=70mm]{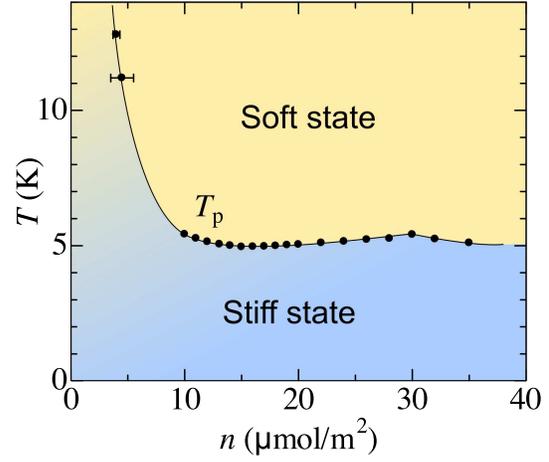}
  \caption{
  (Color online)
  Phase diagram of neon film 
  determined from the dissipation peak temperature $T_\mathrm{p}$. 
  }
  \label{fig:Tp}
 \end{center}
\end{figure}

The height of the dissipation peak $\delta Q^{-1}(T_\mathrm{p})$ grows with increasing $n$ up to $20\ \mathrm{\mu mol/m^2}$. 
At higher coverages, $\delta Q^{-1}(T_\mathrm{p})$ saturates, and the dissipation becomes broadened on the high-temperature side (6--10 K).
The broadening is due to the growth of a second dissipation peak at slightly higher temperatures. 
The coverage dependences of $\delta Q^{-1}(T_\mathrm{p})$ and $T_\mathrm{p}$ are shown in Figs. \ref{fig:n-dfdQ} and \ref{fig:Tp}, respectively. 
We will discuss these behaviors later. 
As seen in Fig. \ref{fig:dfdQinv}(b), the dissipation $\delta Q^{-1}$ becomes negative below about 3 K. 
This apparently negative dissipation 
at low temperatures was also observed in 
the helium film measurements\cite{Makiuchi}.
On the other hand, $\delta Q^{-1}$ above about 15 K is slightly positive. 
The disorder in $\delta Q^{-1}$ at 20 K, seen only at $(4.5\pm 1.0)\ \mathrm{\mu mol/m^2}$, synchronizes with the pressure rise due to the melting of solid neon outside the PG, mentioned in Sect. \ref{sec:NePrep}. 


In addition to the elastic anomaly, we observed a sharp dissipation peak accompanied with an anticrossing behavior of the resonant frequency [shown in the inset in Fig. \ref{fig:dfdQinv}(a)]. 
The sharp dissipation peak occasionally emerged in the measurements of 13--35 $\mathrm{\mu mol/m^2}$. 
The temperature of sharp dissipation peak ranged from 16 to 25 K and had no systematic dependence on the coverage. 
The resonant frequencies on both temperature sides of the anticrossing have little difference. 
From comparisons of the $f$-curves with and without anticrossing [an example is shown in the inset in Fig. \ref{fig:dfdQinv}(a)], the anticrossing was found to reduce $f$ at low temperatures. 
The anticrossing 
clearly originates from a resonant coupling of some vibration mode (e.g., sound wave) in the neon film or porous glass to torsional oscillation, but it is yet to be elucidated. 

The normalized frequency shift $2\delta f(T_\mathrm{min}) / f_0$ at the lowest temperature and the height of the dissipation peak $\delta Q^{-1}(T_\mathrm{p})$ for all examined coverages are shown in Fig. \ref{fig:n-dfdQ}. 
It is clear that $2\delta f (T_\mathrm{min})/f_0$ is linear in $n$ in the coverage range from 4 to 30 $\mathrm{\mu mol/m^2}$, at which it shows a sharp maximum and then decreases. 
In the corrected values (open circles in Fig. \ref{fig:n-dfdQ}), we subtracted the effect of the density change in Eq. (\ref{eq:dfrho}). 
The overall coverage dependence of $2\delta f (T_\mathrm{min})/f_0$ is unchanged between the original and corrected values. 
The height of the dissipation peak $\delta Q^{-1}(T_\mathrm{p})$ is linear in $n$ in a much narrower range of 4 to 18 $\mathrm{\mu mol/m^2}$, beyond which it tends to saturate and shows a shallow minimum at $n=30\ \mathrm{\mu mol/m^2}$, the same coverage as which $\delta f$ has a maximum. 
It is also clear that the linear fittings for both quantities do not pass through the origin and have offset coverages, which are $n_{\mathrm{offset}}=(2.3 \pm 0.5)$ and $(2.7 \pm 0.4)\ \mathrm{\mu mol/m^2}$ for $2\delta f (T_\mathrm{min})/f_0$ and $\delta Q^{-1}(T_\mathrm{p})$, respectively. 
The existence of the offset coverage indicates that the first adsorbed neon atoms up to $n_{\mathrm{offset}}$ do not contribute to the elastic anomaly. 
The coverage $n=18\ \mathrm{\mu mol/m^2}$ at which $\delta Q^{-1}(T_\mathrm{p})$ starts to saturate 
is close to the estimated monolayer coverage of 21.0 $\mathrm{\mu mol/m^2}$. 
It is hypothesized that atoms adsorbed on porous glass do not form distinct layer structures, unlike the case of atoms adsorbed on periodically corrugated substrates such as graphite. 
However, neon atoms firstly fill all adsorption sites adjacent to the substrate.
Then, further atoms are adsorbed on these already adsorbed neon atoms. 
In such a situation, adatoms that are closer to the substrate more prominently influence the dynamical response than adatoms farther from the substrate. 
The saturation of $\delta Q^{-1}(T_\mathrm{p})$ above 18 $\mathrm{\mu mol/m^2}$ suggests that the first monolayer in contact with the glass surface mostly contributes to dissipation, although the second layer still contributes to the elasticity [$2\delta f (T_\mathrm{min})/f_0$].

The normalized frequency $2\delta f (T_\mathrm{min})/f_0$ shows a peak at $n = 30\ \mathrm{\mu mol/m^2}$, above which it rapidly decreases as $n$ increases. 
The coverage of $30\ \mathrm{\mu mol/m^2}$ corresponds roughly to 1.5 layers. 
In helium films, elasticity also becomes maximum at coverages less than the critical one\cite{Makiuchi}.
The ground state of helium films is assumed to be a strongly correlated state with an energy gap in excitation, and the maximum in elasticity 
 originates from the quantum nature of helium films near the QPT point.
The elasticity maximum observed in the neon film also implies correlation between neon atoms, which is caused by the zero-point motion of neon atoms. 


\subsection{Thermal activation model and fitting}
We succeeded\cite{Makiuchi} in quantitatively explaining the elastic anomaly in helium films in terms of a thermal activation model for localized states of helium atoms, which was first proposed by Reppy and co-workers\cite{Tait1979,Crowell1997}, together with the concept of anelastic relaxation\cite{NowickBook}. 
We apply the same procedure to fit the data of neon films. 

The model assumes that adsorbed atoms form two energy bands consisting of localized ground states and spatially extended excited states separated by an energy gap $E$. 
At $T=0$, all atoms occupy the localized states.
At a finite temperature comparable to the energy scale of the gap, atoms are thermally excited from the localized to extended states with thermal relaxation time $\tau=\tau_0 e^{E/k_\mathrm{B}T}$, where $\tau_0^{-1}$ is the attempt frequency. 
In the dynamical experiment at an angular frequency $\omega=2\pi f$, when the relaxation time is larger than the period of oscillation, i.e., $\omega\tau \gg 1$, the atoms are localized and contribute to the elasticity of the substrate.
When $\omega\tau \ll 1$, atoms are frequently excited and localized during the deformation of the substrate; therefore, the elasticity equals the value without the adsorbed atoms.
Dynamic response functions for anelastic relaxation\cite{NowickBook} are
\begin{equation}
 \frac{2\delta f}{f_0} = \frac{\delta G}{G_0}\left[1-\frac{1}{1+(\omega\tau_0e^{E/k_\mathrm{B}T})^2}\right],
 \label{eq:df}
\end{equation}
and
\begin{equation}
 \delta Q^{-1} = \frac{\delta G}{G_0}\frac{\omega\tau_0e^{E/k_\mathrm{B}T}}{1+(\omega\tau_0e^{E/k_\mathrm{B}T})^2},
 \label{eq:dQ}
\end{equation}
where $\delta G$ is the relaxed shear modulus and $G_0$ is the shear modulus of the torsion rod.
At a temperature where $\omega \tau=1$, $\delta Q^{-1}$ has a peak and $2\delta f/f_0$ has a maximal slope.
For the Debye relaxation process with a single-valued activation energy, $2\delta f(T\ll T_\mathrm{p})/f_0 $ is twice as large as $\delta Q^{-1} (T_\mathrm{p})$.

The observation results shown in Fig. \ref{fig:n-dfdQ} give, however, $[2\delta f (T_\mathrm{min})/f_0]/ \delta Q^{-1} (T_\mathrm{p}) >2$ for every coverage.
The linear fits give $[2\delta f (T_\mathrm{min})/f_0]/ \delta Q^{-1} (T_\mathrm{p}) = 9.3 \pm 0.4$. 
This suggests that $E$ has a distribution.
A log-normal distribution, $F(E)= e^{-[\ln (E/\Delta)]^2/2\sigma^2}/\sqrt{2\pi}\sigma E$, where $\Delta$ is the median and $\sigma$ is a dimensionless parameter, is suitable for a symmetric shape of $\delta Q^{-1}$ at $T_\mathrm{p}$ against the log of the temperature. 
Finally, the response function in complex form is\cite{Makiuchi}
\begin{equation}
 \frac{2\delta f}{f_0} + i \delta Q^{-1} = \frac{\delta G}{G_0}\left[1-\int_0^\infty \frac{F(E)}{1+i\omega\tau_0e^{E/k_\mathrm{B}T}} dE\right].
 \label{eq:df+idQinv}
\end{equation}
Hereafter, we regard the median $\Delta$ as the energy gap.

\begin{figure}
 \begin{center}
  \centering
  \includegraphics[width=70mm]{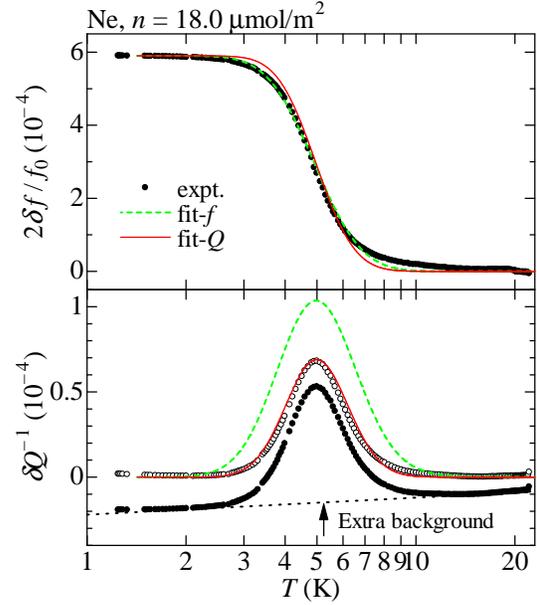}
  \caption{
  (Color online)
  Experimental data with $n=18\ \mathrm{\mu mol/m^2}$ and results of fittings using Eq. (\ref{eq:df+idQinv}).
  The experimental data of $2\delta f/f_0$ are vertically shifted. 
  Closed circles of $\delta Q^{-1}$ data were determined using Eq. (\ref{eq:deltaQinv}), and open circles are data from which an extra background proportional to $\log(T)$ was subtracted.
  The closed circles and the dotted line are vertically shifted by $-0.1\times 10^{-4}$.
  Dashed curve (fit-$f$) matches $2\delta f/f_0$ but does not $\delta Q^{-1}$.
  Solid curve (fit-$Q$) was found to fit better with both $2\delta f/f_0$ and $\delta Q^{-1}$. 
  The fitting parameters employed in fit-$f$ (fit-$Q$) were $\delta G/G_0 = 5.9\times 10^{-4}$ (the same), $\tau_0 = 4\times 10^{-10}$ ($4\times 10^{-15}$) s, $\Delta /k_\mathrm{B}$ = 64 (122) K, and $\sigma = 0.25$ (0.21).
  }
  \label{fig:fit}
 \end{center}
\end{figure}

Figure \ref{fig:fit} shows fittings of equations with $n=18.0\ \mathrm{\mu mol/m^2}$ data.
The dissipation $\delta Q^{-1}$ has an extra background, which is proportional to $\log(T)$, and we subtracted it from $\delta Q^{-1}$ obtained from Eq. (\ref{eq:deltaQinv}) before the fittings. 

In fit-$f$, the lower-temperature shoulder of $2\delta f/f_0$ matches well with the data.
The position and the symmetry of the dissipation peak of fit-$f$ also agree with the data. 
The values of the inverse attempt frequency $\tau_0 =0.4$ ns and the ratio $\Delta/k_\mathrm{B}T_\mathrm{p} = 13$ are common to helium films\cite{Makiuchi}. 
The width and height of $\delta Q^{-1}$ in fit-$f$, however, disagree with the experimental results.

We attempt, in fit-$Q$, to match a fitting curve with $\delta Q^{-1}$ data. 
The width and height of $\delta Q^{-1}$ are reduced by a factor of about 0.7 (compare fit-$f$ and fit-$Q$ in Fig. \ref{fig:fit}) upon lowering $\tau_0$ by several orders of magnitude.
We obtain $\tau_0=4$ fs and $\Delta/k_\mathrm{B} = 122$ K ($\Delta /k_\mathrm{B}T_\mathrm{p} = 24$).
From the fit-$Q$ of other coverages, we get $\tau_0$ values of femtosecond order or much smaller. 
Fit-$Q$ agrees better than fit-$f$ with the experimental data in the entire temperature range. 
The obtained $\tau_0$ is, however, too small to clarify the physical meaning.
It is unlikely that neon has such a larger attempt frequency $\tau_0^{-1}$ of several orders of magnitude than helium. 
If we take the distribution of the energy gap to be roughly equal to the median $\Delta $, the product $ \tau_0\Delta \sim 10^{-36}\ \mathrm{Js}$ is smaller than the Planck constant $h$, and this violates the time--energy uncertainty principle.
Further physical insight is necessary to solve this problem.


\subsection{Phase diagram determined by $T_\mathrm{p}$} 
We propose a phase diagram based on the coverage dependence of the dissipation peak temperature $T_\mathrm{p}$. 
It is shown in Fig. \ref{fig:Tp}. 
The dissipation peak temperature strongly depends on the coverage below $n=$ 10 $\mathrm{\mu mol/m^2}$, but above this coverage, it becomes almost constant. 
There is a cusp in the $n$--$T_\mathrm{p}$ curve at $n=30\ \mathrm{\mu mol/m^2}$, the same coverage at which both $\delta f(T_\mathrm{min})$ and $\delta Q^{-1}$ show the features (in Sect. \ref{sec:anomaly}).

One may regard $T_\mathrm{p}$ as the temperature that distinguishes the two states of a neon film, a stiff state and a soft state. 
If each of the states corresponds to different phases, there exist only two phases in neon films adsorbed on PG. 
This contrasts with the five phases of a submonolayer neon film on graphite\cite{Huff1976,Rapp1981}. 
A similar situation has been found in helium: There are two phases at $T=0$, a gapped solid phase and a mobile (superfluid) phase, in $^4$He and $^3$He films on porous Vycor glass and on Gelsil\cite{Crowell1997,Makiuchi}, while helium on graphite shows a number of phases depending on the coverage\cite{Greywall1990,Greywall1993}. 
The disordered potential provided by porous glass will make the phase diagram more simple. 

\section{Discussion}
At each coverage, the elastic anomaly of neon films is qualitatively identical to that of helium films. 
There exists, however, an important difference. 
In each phase diagram of $^4$He and $^3$He films\cite{Makiuchi}, $T_\mathrm{p}$ approaches 0 K and the vanishing point is the critical coverage, $n_\mathrm{c}\simeq 20\ \mathrm{\mu mol/m^2}$. 
Hence, the elastic anomaly of helium films is linked to the gapped solid phase on one side of the QPT, which is indeed a superfluid--insulator QPT in the case of $^4$He. 
On the other hand, $T_\mathrm{p}$ of neon does not decrease to below 5 K. 
This indicates that a neon film on PG is close to a classical system, which neither shows the QPT nor becomes superfluid. 
The elastic anomaly is therefore not limited to the accompanying phenomenon of the QPT, but rather originates from the energy structure of adsorbed films. 


As the elastic anomaly is not sharp but gradual, the crossover from the stiff state to the soft state does not indicate a first-order phase transition or Kosterlitz--Thouless-type melting, contrary to the case of bulk and monolayer neon on graphite\cite{Huff1976,Rapp1981}. 
One may naturally expect neon atoms on disordered and porous substrates to form an amorphous film. 
The amorphous state is consistent with the fact that layer-by-layer growth of a neon film was not observed on disordered substrates\cite{Huber1995}, while neon films on graphite show clear layer-by-layer growth and sharp signatures in the heat capacity\cite{Rapp1981}. 

To elucidate the elastic anomaly, it is important to evaluate the stiffness of our neon films. 
The normalized resonant frequency shift in the stiff state, $2\delta f(T_\mathrm{min})/f_0$ becomes maximum at $n=30\ \mathrm{\mu mol/m^2}$, as shown in Fig. \ref{fig:n-dfdQ}.
At this coverage, the effective shear modulus $\delta G_\mathrm{g}$ is calculated to be 40.1 MPa using $2\delta f(T_\mathrm{min})/f_0=1.07\times 10^{-3}$, $G_\mathrm{g0}=7.38$ GPa\cite{Makiuchi}, and Eq. (\ref{eq:dfG}).
As $2\delta f(T)/f_0$ does not change in the range 1.2--3 K and all neon atoms solidify at $T=0$, the stiff state is a solidlike state.
The effective shear modulus is smaller than the shear modulus of bulk solid neon, $G_\mathrm{Ne} = 600$ MPa\cite{White1998}, but this is reasonable because the neon film partially fills the pores and the porous structure of the neon film results in a small effective shear modulus\cite{Pabst2006}.
On the other hand, in the soft state, $2\delta f(22\ \mathrm{K})/f_0<3\times 10^{-5}$, i.e., $\delta G_\mathrm{g}<$  1 MPa, for every coverage.
Furthermore, $2\delta f(22\ \mathrm{K})/f_0$ does not change systematically with increasing $n$, which suggests that additional neon atoms in the soft state do not contribute to the elasticity.
The large difference in $\delta G_\mathrm{g}$ between the stiff and soft states suggests that the soft state is fluidlike.
The fluidlike state appearing at temperatures much lower than the bulk triple-point temperature 24.6 K might be realized by strong supercooling of thin neon films in nanoscale confinement.

Finally, we point out that the $T$ dependences of frequency and dissipation data (shown in Fig. \ref{fig:dfdQinv}) bear a strong resemblance to the elastic behavior seen in glass transitions. 
In glass transitions of polymers and amorphous materials, the elasticity or the velocity of sound gradually increases accompanying a peak in dissipation or attenuation as the system passes the dynamic glass transition temperature $T_\mathrm{g}^\mathrm{dyn}$\cite{Kruger2007}. 
Our dissipation peak temperature $T_\mathrm{p}$ may correspond to $T_\mathrm{g}^\mathrm{dyn}$ or may be related to it. 
In order to clarify the aspects of the glass transition in the elastic anomaly of neon films, experiments focused on glassy properties such as the dependence on cooling velocity will be helpful. 


\section{Conclusions}
We found that neon films on porous glass exhibit an elastic anomaly, in which the elasticity increases with excess dissipation. 
The elastic anomaly is qualitatively identical to that of helium films.
The elasticity at the lowest temperature and the height of the dissipation peak are proportional to the coverage with an offset up to certain coverages, and both showed kinks at the same coverage of 30 $\mathrm{\mu mol/m^2}$.
The data were fitted by a thermal activation model that successfully explains the elastic anomaly of helium films. 
In contrast to helium films, the dissipation peak temperature becomes a constant of about 5 K, suggesting that neon films behave as a classical system without a quantum phase transition with changing coverage. 
The crossover from the stiff state to the soft state bears a resemblance to the dynamic glass transition.

Our present work showed that the elastic anomaly is observed not only in helium films but also in neon films. 
We recently observed similar elastic anomaly in hydrogen films\cite{hydrogen}. 
Therefore, the elastic anomaly is probably a universal phenomenon of atomic or molecular films adsorbed on disordered substrates. 
Future investigations using other adsorbed molecules will unveil the general nature of adsorbed states and their dynamics. 

\begin{acknowledgment}
 This work was supported by JSPS KAKENHI Grant No. JP17H02925.
 TM was supported by Grant-in-Aid for JSPS Research Fellow JP18J13209.
\end{acknowledgment}


\begin{thebibliography}{19}
\bibitem{hydrogen}We performed elastic measurements for adsorbed films of H$_2$, HD, and D$_2$, and found similar elastic anomalies in all these molecular films. The results will be published elsewhere.
\end{thebibliography}
\end{document}